\documentclass[fleqn,twoside]{article}
\usepackage{amssymb}
\usepackage{espcrc2}

\usepackage{graphicx}
\usepackage[figuresright]{rotating}

\newcommand{\AmS}{{\protect\the\textfont2
  A\kern-.1667em\lower.5ex\hbox{M}\kern-.125emS}}

\newcommand{\be}{\begin{equation}}
\newcommand{\ee}{\end{equation}}
\newcommand{\bea}{\begin{eqnarray}}
\newcommand{\eea}{\end{eqnarray}}
\newcommand{\Id}{\mathbf{1}}
\newcommand{\vx}{\mathbf{x}}
\newcommand{\vy}{\mathbf{y}}

\newcommand{\cO}{\mathcal{O}}
\newcommand{\cK}{\mathcal{K}}

\title{\vspace{-4.0cm}
       \rightline{\normalsize ROM2F/2001/34}
       \vspace{-0.1cm}
       \rightline{\normalsize MS-TP-01-9}
       \vspace{-0.1cm}
       \rightline{\normalsize CERN-TH/2001-276}
       \vspace{-0.1cm}
       \rightline{\normalsize October 2001}
       \vspace{2.0cm}
       $K^0$-$\bar{K}^0$ mixing from the Schr\"odinger functional and twisted mass QCD\thanks
       {Based on a talk by C.~Pena at the International Symposium of Lattice Field Theory,
       August 19-24 2001, Berlin, Germany.}
      }

\author{M.~Guagnelli\address[ToV]{INFN Sezione Roma II,
                                  c/o Dipartimento di Fisica, Universit\`a di Roma ``Tor Vergata'',\\
                                  Via della Ricerca Scientifica 1, I-00133 Rome, Italy.},
        J.~Heitger\address[Muenster]{Westf\"alische Wilhelms-Universit\"at M\"unster,
                                     Institut f\"ur Theoretische Physik,\\
                                     Wilhelm-Klemm-Stra\ss e 9, D-48149 M\"unster, Germany.
                                     },
        C.~Pena\addressmark[ToV],
        S.~Sint\address[CERN]{CERN, Theory Division, CH-1211 Geneva 23, Switzerland.} and
        A.~Vladikas\addressmark[ToV]
       }

\begin{document}

\begin{abstract}
We describe how the Schr\"odinger functional and twisted mass QCD can be used to
compute the kaon $B$-parameter $B_K$ on the lattice with Wilson fermions. This new approach
is expected to reduce the systematic uncertainties on $B_K$ through a better control
of its (nonperturbative) renormalisation. Preliminary results for the bare matrix element in a
physical volume of $1.5^3 \times 4.5 \mbox{~fm}^4$ are presented. The renormalisation of the
matrix element in a Schr\"odinger functional scheme is also discussed.
\end{abstract}
\vspace{0mm}

\maketitle

\section{INTRODUCTION}
Out of the physical quantities related to CP violation in the Standard Model, the parameter
$\varepsilon$ accounting for indirect CP violation in the kaon system is the one with the simplest
non-perturbative contribution. It is given  by the kaon $B$-parameter $B_K$, defined as:
\be
\label{def_BK}
B_K = \frac{\langle\bar{K}^0 \vert O^{\Delta S=2} \vert K^0\rangle}{\frac{8}{3}F_K^2 m_K^2} \ ,
\ee
where $O^{\Delta S=2}$ is the $\Delta S=2$ operator of the effective weak Hamiltonian:
\be
\label{def_deltas2}
O^{\Delta S=2} = (\bar{s}\gamma_{\mu}^Ld)(\bar{s}\gamma_{\mu}^Ld) \ ,
\ee
and $\gamma_{\mu}^L = \scriptsize{\frac{1}{2}}\gamma_{\mu}(\Id-\gamma_5)$. In an obvious
notation, it is convenient to split $O^{\Delta S=2}$ in parity-even and parity-odd parts:
\be
\label{split_O}
O^{\Delta S=2} = O^{\Delta S=2}_{VV+AA}-O^{\Delta S=2}_{VA+AV} \ .
\ee
Since parity is a QCD symmetry, only the parity-even part $O^{\Delta S=2}_{VV+AA}$ will
yield a nonvanishing $K^0$-$\bar{K}^0$ matrix element.

In spite of the absence of both power-divergent subtractions and final state interactions,
the lattice computation of $B_K$ is still a very difficult problem. Besides the systematic error 
arising from the quenched approximation, the breaking of chirality (due to the use of Wilson
fermions) leads to a mixing under renormalisation of $O^{\Delta S=2}$ with four other operators
of dimension six. It is now understood that the corresponding
mixing coefficients, as well as the overall, scale-dependent, renormalisation constant,
require an accurate nonperturbative computation. This has been dealt with by using
Ward identities and RI-MOM renormalisation~\cite{BKRoma,BKJap}, but the result still reflects large numerical
uncertainties. On the other hand, the use of lattice actions with better chiral
properties has yielded results which have not manifested clear agreement
with those obtained with Wilson fermions. We refer the reader to~\cite{MERev} for reviews on the subject
and references.

To improve on this situation, we propose a new computation of $B_K$ with Wilson fermions. 
It adopts the Schr\"odinger functional (SF)~\cite{SF} for the computation of
scale-dependent renormalisation constants, and a twisted mass QCD (tmQCD) action~\cite{tmQCD1}, in order
to avoid the mixing of $O^{\Delta S=2}$ with other operators under renormalisation. An alternative 
proposal for the elimination of this mixing has been presented in~\cite{altro}.

\section{tmQCD FOR $B_K$}
Twisted mass QCD was proposed as a means of avoiding the problem of exceptional
configurations, allowing lattice QCD computations with light quark masses closer
to the chiral limit~\cite{tmQCD1}. For the computation of $B_K$ we will use an action with a
twisted doublet of light quarks $\psi=(u,d)$ and a standard (untwisted) strange quark $s$:
\setlength{\arraycolsep=1pt
\bea
\nonumber
S &=& a^4\sum_x\big[\bar{\psi}(x)\left(D+m_l+i\mu_l\gamma_5\tau^3\right)\psi(x) \\
\label{action}
&& \relax{\kern+10mm} + \bar{s}(x)\left(D+m_s\right)s(x)\big] \ ,
\eea
}
where $\tau^3$ acts on isospin indices and $\mu_l$ is the ``twisted mass''.
In the classical theory at the continuum limit, the standard QCD action is recovered
by changing fermionic variables through an axial rotation of the form
$\psi\to\exp\{-i\alpha\gamma_5\tau^3/2\}\psi$, where $\tan(\alpha)=\mu_l/m_l$.
This rotation leaves invariant the combination $M_l=\sqrt{m_l^2+\mu_l^2}$,
which can therefore be regarded as the physical bare light quark mass;
on the other hand, it induces a mapping between composite operators
in the two theories. It has been shown in~\cite{tmQCD1} that this classical
equivalence carries over to appropriately renormalised correlation functions
in the quantum theory. The twisting angle is then given by
$\tan(\alpha)=\mu_{l,R}/m_{l,R}$, where the masses are renormalised.

By fixing renormalised masses so that $\alpha=\pi/2$, the mapping
between operators induced by the corresponding axial rotation leads
to the relation:
\setlength{\arraycolsep=1pt
\bea
\nonumber
\langle\bar{K}^0\vert O^{\Delta S=2}_{VA+AV}\vert K^0\rangle_{\mbox{\scriptsize{tmQCD}}}^{\alpha=\pi/2} &=&\\
\nonumber
&&\relax{\kern-15mm} = i \langle\bar{K}^0\vert O^{\Delta S=2}_{VV+AA}\vert K^0\rangle_{\mbox{\scriptsize{QCD}}} \\
\label{twist_BK}
&&\relax{\kern-15mm} = i\langle\bar{K}^0\vert O^{\Delta S=2}\vert K^0\rangle_{\mbox{\scriptsize{QCD}}}
\eea
}
(the above is to be understood as valid for renormalised matrix elements). Parity violation
at nonzero twisted mass $\mu_l$ ensures that the tmQCD matrix element does not vanish.
Computing the lhs of Eq.~(\ref{twist_BK}), rather than the
rhs, is advantageous, as the parity-odd operator $O^{\Delta S=2}_{VA+AV}$ is protected by
$\mathcal{CPS}$ symmetry from mixing with other operators, and hence it renormalises
in a purely multiplicative fashion~\cite{Bernard-EPJ}. This is clearly true if renormalisation is
carried out in the chiral limit, where tmQCD and QCD (and their symmetries) coincide.

\begin{figure}[htb]
\vspace{60mm}
\includegraphics{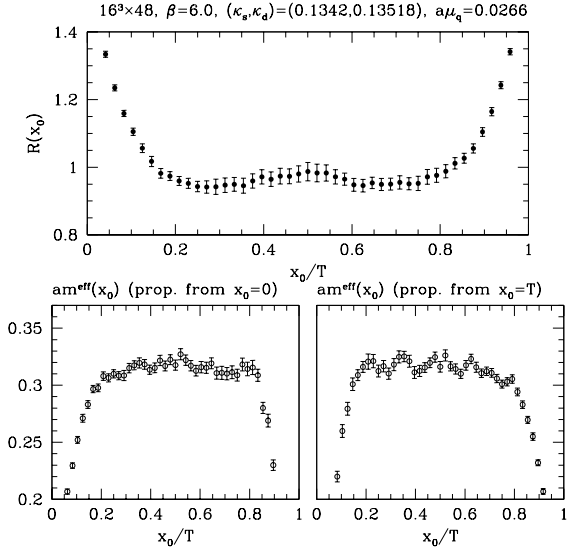}   
\caption{Plateaux for $B_K$ and effective kaon masses, from 250 configurations, for
a $16^3 \times 48$ lattice at $\beta=6.0$.}
\label{fig:BK}
\end{figure}

We show the feasibility of a direct computation of the matrix element of $O^{\Delta S=2}_{VA+AV}$
with the tmQCD action and SF boundary conditions by presenting preliminary results for the
bare value of $B_K$, obtained on a $16^3 \times 48$ lattice at $\beta = 6.0$. This corresponds
approximately to a physical volume of $1.5^3 \times 4.5 \mbox{~fm}^4$. In the computation we
have kept the $s$ and $d$ masses degenerate and close to the physical $s$ quark mass,
as done in previous computations of $B_K$. We show in Fig.~\ref{fig:BK} the
ratio of correlations:
\be
\label{ratio4BK}
R(x_0) = \frac{-i\langle\cO^\prime_{ds}O^{\Delta S=2}_{VA+AV}(x)\cO_{ds}\rangle}
              {\frac{8}{3}\langle\cO^\prime_{ds}\cK(x)\rangle\langle\cK(x)\cO_{ds}\rangle} \ ,
\ee
where $\cK=-\scriptsize{\frac{i}{\sqrt{2}}}\bar{s}\gamma_0(\Id+i\gamma_5)d$ is the operator obtained by axially rotating
the QCD $A_0$ current with  $\alpha=\pi/2$, and $\cO,\cO^\prime$
are the standard SF pseudoscalar sources:
\bea
\label{def_source_1}
\cO_{f_1f_2} &=& L^{-3}\sum_{\vx\vy}\bar{\zeta}_{f_1}(\vx)\gamma_5\zeta_{f_2}(\vy) \ , \\
\label{def_source_2}
\cO^\prime_{f_1f_2} &=& L^{-3}\sum_{\vx\vy}\bar{\zeta}'_{f_1}(\vx)\gamma_5\zeta'_{f_2}(\vy) \ ,
\eea
where $f_1,f_2$ label two distinct flavours.
The central plateau of $R$ corresponds to the \underline{bare}
value of $B_K$. As expected, it appears
in the time interval $0.35 \lesssim x_0/T \lesssim 0.65$, where the single kaon states coming from Euclidean
times $0$ and $T$ are isolated (as shown by the effective mass plots of Fig.~1).
Fitting the $R$-plateau to a constant we get $B_K(a^{-1} \sim 2 \mbox{~GeV}) = 0.96(2)$.

\section{SF RENORMALISATION}
The renormalisation constant of the matrix element at renormalisation scale
$\mu=L^{-1}$ can be evaluated on symmetric lattices of physical volume $L^4$. The SF renormalisation
condition, which is imposed at the chiral limit $m_{l,R}=m_{s,R}=\mu_{l,R}=0$, is then:
\be
\label{renorm_cond}
Z_{VA+AV}(g_0,L)\frac{F(L/2)}{f_1^{3/2}} = \left.\frac{F(L/2)}{f_1^{3/2}}\right\vert_{g_0^2=0} \ ,
\ee
where $f_1$ is a wave function renormalisation factor, defined as:
\setlength{\arraycolsep=1pt
\be
\label{def_f1}
f_1 = -\frac{1}{2}\langle \cO^\prime_{sd}\cO_{ds}\rangle \ ,
\ee
}
and $F$ is a correlator involving $O^{\Delta S=2}_{VA+AV}$, chosen so as not to vanish
in the chiral limit. We opt for the simple choice:
\be
\label{rc_corr}
F(x_0) = L^6\langle\cO^\prime_{us} O^{\Delta S=2}_{VA+AV}(x) \cO_{ds}\cO_{du}\rangle \ .
\ee

Once $Z_{VA+AV}$ has been defined at given values of the bare coupling $g_0$ and of the
renormalisation scale $\mu=L^{-1}$, a step scaling function that describes the (nonperturbative)
running with $\mu$ can be defined at fixed value $u=\bar{g}^2(L)$ of the renormalised coupling as:
\be
\label{def_ssf}
\sigma_{VA+AV}(u) = \lim_{a \to 0}\left.\frac{Z_{VA+AV}(g_0,2L)}{Z_{VA+AV}(g_0,L)}\right\vert_{\bar{g}^2(L)=u}
\relax{\kern-3mm} \ .
\ee
Having computed $\sigma_{VA+AV}$ for a number of values of $u$, it is then possible to
run the renormalised matrix element to large scales $\mu=L^{-1}$, at which the perturbative
matching to other renormalisation schemes, such as $\overline{\mbox{MS}}$, can be performed.

By computing various $Z$ constants, obtained from different sources $\cO, \cO^\prime$
(i.e. different renormalisation conditions), we hope to get an idea of the size of
$O(a)$ effects, which arise due to the use of the unimproved four-quark operator.

\section{CONCLUSIONS}
We have described how the SF and tmQCD can be used to achieve a better control of the renormalisation properties
of the kaon $B$-parameter $B_K$. A first feasibility study for the bare matrix
element giving $B_K$ in tmQCD has been presented, and the strategy for the SF renormalisation
discussed.

We mention that the same techniques can be extended to all $\Delta S = 2$ operators of interest,
thus avoiding the spurious mixing with operators of the same dimension
induced by the use of Wilson fermions. The computation of the step scaling
functions for the complete basis of such operators in already under way.

\section*{ACKNOWLEDGEMENTS}
We wish to thank D.~Be\'cirevi\'c, K.~Jansen, G.~Martinelli and R.~Sommer for useful discussions. This work was supported
in part by the European Community's Human Potential Programme under contract HPRN-CT-2000-00145,
Hadrons/Lattice QCD. C.P. acknowledges the financial support provided through this contract.

%%%%%%%%%%%%%%%%%%%%%%%%%%%%%%%%%%%%%%%%%%%%%%%%%%%%%%%%%%%%%%%%%%%%%%%%%%%%%%%%%%%%%%%%%%%%%

\end{document}